\documentclass[prd,twocolumn,aps,tightenlines,preprintnumbers,notitlepage,
superscriptaddress,floatfix]{revtex4-2}
\usepackage[dvipsnames]{xcolor}
\usepackage{color}
\usepackage{tikz}
\usepackage[caption=false]{subfig}
\usepackage{multirow}
\usepackage{siunitx}
\usepackage[export]{adjustbox}
\usepackage{graphicx}
\usepackage{dcolumn}
\usepackage{bm}
\usepackage[normalem]{ulem}
\usepackage{epstopdf}
\usepackage{tabularx}
\usepackage{suffix}
\usepackage{mathtools}
\usepackage{graphicx}
\usepackage{dcolumn}
\usepackage{bm}
\usepackage[normalem]{ulem}
\usepackage{xcolor}
\usepackage{epstopdf}
\usepackage{lipsum}
\usepackage{tabularx} 
\newcolumntype{L}{>{\raggedright\arraybackslash}X} 
\newcolumntype{C}{>{\centering\arraybackslash}X}   
\usepackage{hyperref}

\definecolor{TurkishBlue}{HTML}{144893}
\definecolor{TurkishBlue2}{HTML}{1985B6}

\hypersetup{colorlinks=true,citecolor=TurkishBlue2,linkcolor=TurkishBlue2,urlcolor=TurkishBlue2}

\graphicspath{ {figures/} }

\definecolor{mypink}{HTML}{C34D85}

\def\bk{\boldsymbol{k}}

\def\bL{\boldsymbol{L}}
\def\br{\boldsymbol{r}}
\def\bq{\boldsymbol{q}}
\newcommand{\bn}{\hat{ \mathbf{n}}}
\newcommand{\hatv}{\hat{v}}

\newcommand{\bl}{{\boldsymbol{\ell}}}
\newcommand{\dd}{{\rm d}}
\newcommand{\td}[1]{{\tilde{#1}}}

\def\ksz{{\rm kSZ}}
\def\ion{{\rm ion}}

\newcommand{\DdeltaC}{(2\pi)^3\delta^3}
\newcommand{\DdeltaS}{(2\pi)^2\delta^2}

\newcommand{\ellInt}[2][]{\frac{\dd^2\bl_{#2}^{#1}}{(2\pi)^2}}
\newcommand{\kInt}[2][]{\frac{\dd^3\bk_{#2}^{#1}}{(2\pi)^3}}

\def\VEV#1{{\left\langle #1 \right \rangle}}
\def\lr#1#2#3{{\left#1 #2 \right#3}}

\newcommand{\be}{\begin{eqnarray}}

\newcommand{\ee}{\end{eqnarray}}

\newcommand{\aj}{Astron. J.}

\newcommand{\mnras}{Mon. Not. R. Astron. Soc.}

\newcommand{\Pion}{P^{\rm ion}_{ee}}

\newcommand{\jhu}{William H.~Miller III Department of Physics and Astronomy, Johns Hopkins University, 3400 N Charles St, Baltimore, MD 21218, USA}

\newcommand{\perimeter}{Perimeter Institute for Theoretical Physics, 31 Caroline St N, Waterloo, ON N2L 2Y5, Canada}

\begin{document}

\title{A Novel kinetic Sunyaev-Zel'dovich Estimator for Electron-Electron Correlations}
\author{Neha~Anil~Kumar}
\affiliation{\jhu}

\author{Mesut~\c{C}al{\i}\c{s}kan}
\affiliation{\jhu}

\author{Selim~C.~Hotinli}
\affiliation{\perimeter}

\author{Kendrick~Smith}
\affiliation{\perimeter}

\author{Marc~Kamionkowski}
\affiliation{\jhu}

\date{}


\begin{abstract}
\noindent Recent advancements in small-scale observations of the cosmic microwave background (CMB) have provided a unique opportunity to characterize the distribution of baryons in the outskirts of galaxies via stacking-based analyses of the kinetic Sunyaev-Zel'dovich (kSZ) effect. 
Such measurements, mathematically equivalent to probing the galaxy-electron cross-correlation, have revealed that gas is more extended than dark matter and that the strength of baryonic feedback may vary with halo mass and redshift. 
However, because these analyses are conditioned on galaxy positions, deriving a host-independent description of the baryon distribution depends on uncertain galaxy-halo modeling on small scales.
In this work, we present a novel kSZ$\times$galaxy four-point estimator that directly probes the full ionized electron field, extending beyond the gas traced by luminous galaxies.
This method exploits large-scale velocity reconstruction from galaxy surveys to characterize the electron distribution \textit{unbiased} by small-scale galaxy clustering.
We forecast that the proposed signal can be measured with a signal-to-noise ratio of $\sim8$ ($\sim31$) for a configuration corresponding to Atacama Cosmology Telescope DR6 (Simons Observatory) CMB data combined with spectroscopic galaxy samples from DESI.
This approach will enable the first tomographic measurements of the electron auto-power spectrum, providing new constraints on feedback-driven redistribution of baryons and its role in shaping cosmic structure.
\end{abstract}

\maketitle


\textit{Introduction}---Baryonic feedback from galaxies, driven by processes such as supernova outflows and active galactic nuclei, plays a central role in shaping the distribution of ionized gas in halos and their surroundings~\citep{2001ApJ...552..473D,Fukugita:2004ee,2006ApJ...650..560C}.
This feedback leaves measurable imprints on the ionized-electron distribution that can be directly probed with observations of cosmic microwave background (CMB) secondary signals induced by interactions of CMB photons with the free-electron distribution along the line of sight.

A prominent probe in this regard is the kinetic Sunyaev-Zel'dovich (kSZ) effect~\citep{Sunyaev:1972eq, Sunyaev:1980nv, 1991ApJ...372...21R, Birkinshaw:1998qp}, a Doppler-induced temperature anisotropy arising from the scattering of CMB photons off ionized electrons with bulk velocities relative to the CMB-rest frame. 
Measurements of the~ionized gas distribution using the kSZ signal have primarily relied on stacking analyses centered on the positions of luminous-red galaxies~\citep{DESI:2022gle} and bright-galaxy sample targets~\citep{Hahn:2022dnf}, which are dominated by halos in the galaxy-group mass range.
Such measurements recover the small-scale electron density profiles conditioned on galaxy locations, or equivalently, the small-scale galaxy-electron power spectrum $P_{ge}(k,z)$~\citep{Smith:2018bpn}. 
The inferred electron distributions are, therefore, tied to uncertain galaxy-halo modeling on small scales and best characterize gas distributions surrounding the selected host samples.

Recently, advances in CMB and large-scale structure observations, provided by the Atacama Cosmology Telescope (ACT)~\citep{ACT:2020gnv, ACT:2023kun, ACT:2025xdm, ACT:2023kun} and the Dark Energy Spectroscopic Instrument (DESI)~\citep{DESI:2022xcl,DESI:2013agm,DESI:2016fyo,2019AJ....157..168D,Zhou:2023gji}, have allowed for characterization of the ionized-gas density in the outskirts of galaxies using this technique~\citep[e.g.,][]{Hadzhiyska:2025mvt, RiedGuachalla:2025byu, ACT:2025llb, Hadzhiyska:2024ecq, Hadzhiyska:2024qsl, Hadzhiyska:2023cjj}. Strikingly, these observations challenge the predictions of some state-of-the-art hydrodynamical simulations~\citep[e.g.,][]{Schaye:2023jqv, Dave:2019yyq, Nelson:2018uso} that have feedback prescriptions calibrated on earlier generations of cluster X-ray data. 
These measurements suggest that significant fractions of baryons have been displaced to larger radii than previously anticipated, and that feedback effects may vary non-trivially with the mass and epoch of the halo~\citep[e.g.,][]{Sunseri:2025hhj,Bigwood:2025ism,Lucie-Smith:2025hgj,Kovac:2025zqy, McCarthy:2024tvp,Siegel:2025frt}. 
Such departures are compelling because they indicate that current feedback models may be missing key physics that has important implications on our understanding of galaxy formation and evolution. 
Baryonic feedback also complicates the interpretation of cosmological observables by altering the matter and lensing power spectra on non-linear scales~\citep{2011MNRAS.415.3649V, Chisari:2019tus}, as well as biasing the reconstructed velocities from kSZ tomography~\citep{Hotinli:2025tul,Lai:2025qdw,Lague:2024czc,Krywonos:2024mpb,Bloch:2024kpn,Tishue:2025zdw,Kumar:2022bly,Hotinli:2022jna,AnilKumar:2022flx,Cayuso:2021ljq,Hotinli:2020csk,Hotinli:2019wdp,Munchmeyer:2018eey}. These modeling uncertainties therefore make the characterization of feedback essential not only for astrophysics but also for precision cosmology.

A key advantage of the kSZ effect is its sensitivity to the momentum of \textit{all} ionized electrons.
Therefore, it has the potential to constrain the electron density field, including feedback-driven structure, in regions not traced by luminous galaxies. 
However, stacking analyses rely on luminous galaxy positions, which limits direct inferences about the electron density field to regions surrounding these massive hosts.
In this work, we present a novel kSZ$\times$galaxy cross-correlation that accesses the \textit{full} electron auto-power spectrum $P_{ee}^\ion(k,z)$, thereby providing a complementary, host-independent constraint on small-scale baryon distributions.

\textit{Overview of our method}---The method proposed in this work exploits the fact that large-scale variations in the locally measured (high-$\ell$) kSZ power arise from small-scale electron distributions being modulated by long-wavelength perturbations in the cosmological radial-velocity-\textit{squared} field. Therefore, by cross-correlating the kSZ-squared map with a 3D template of the velocity-squared field reconstructed from galaxies, we obtain a tomographic measurement of the cross-spectrum. 
\textit{Crucially}, because the velocity-squared template is derived from galaxies on large scales, the shape of the resulting cross-spectrum depends only on standard cosmology, while its amplitude is directly sensitive to the electron distribution at that redshift. 
In other words, this cross-correlation leverages galaxy-based velocity-squared reconstruction in the linear regime to tomographically extract the small-scale electron clustering information imprinted in the kSZ, \textit{unbiased} by assumptions about the small-scale distribution of galaxies.

\textit{Signal in 3D Box Formalism}---In this work, we approximate the kSZ power spectrum using the following integral~\citep{Shaw:2011sy}:
\begin{equation}
    C_{\ell}^{\ksz} = \int \dd z\ Q(z) \VEV{v_r(z)^2}P_{ee}^{\ion}(\ell/\chi(z), z)\,,
    \label{eq: kSZ_expression}
\end{equation}
where $\VEV{v_r(z)^2} = \VEV{v(z)^2}/3$ is the mean-squared radial velocity, and $\chi(z)$ is the radial comoving distance to redshift $z$. 
The radial weight function $Q(z)$ is defined as:
\begin{equation}
    Q(z) \equiv \bar{T}_{\rm CMB}^2\frac{H(z)}{\chi^2(z)}\left(\frac{\dd\bar{\tau}}{\dd z}\right)^2e^{-2\bar{\tau}(z)}\,,
\end{equation}
where $\bar{T}_{\rm CMB}$ is the average temperature of the CMB, $H(z)$ is the Hubble parameter, and $\bar{\tau}(z)$ is the average optical depth of the scattering process out to $z$ [Eq.~\eqref{eq: tau_bar_eq}].

Next, to quantify large-scale fluctuations in the locally measured kSZ power, we follow the formalism detailed in Ref.~\cite{Smith:2016lnt}. Let $K(\bn) \equiv T_S(\bn)^2$ be the small-scale power near sky-location $\bn$. 
Here, $T_S(\bn)$ represents a high-pass filtered CMB map, defined in Fourier space using an optimal filter $W_S(\bl)$, i.e., $T_S(\bl) \equiv W_S(\bl)T(\bl)$. 
Moreover, let $\bar{K}$ then represent the all-sky average of this field (i.e., $\bar{K}\equiv \VEV{K(\bn)}$). 
The $\eta$-model from Ref.~\cite{Smith:2016lnt} leverages the fact that large-scale modulations in the kSZ power can be attributed to electron distributions (`patchy' on small scales) experiencing long wavelength radial-velocity perturbations, giving the following ansatz for the kSZ contribution to the $K(\bn)$ field sourced by a 3D box centered at redshift $z_*$\,:
\begin{equation}
    K(\bn) = \frac{\dd\bar{K}}{\dd \chi}\Bigg|_{\chi_*}\int_0^{L_0}\dd r\, \eta(\chi_*\bn\,, r\,\hat{\br})\,,
    \label{eq: K_field_expression}
\end{equation}
where $\chi_* \equiv \chi(z_*)$, $L_0$ is the comoving side length of the box, and $\eta(\bn,z) \equiv v_r(\bn,z)^2/\VEV{v_r(z)^2}$.
In this model, the dependence of the $K(\bn)$ field on small-scale physics is encapsulated in the `amplitude' pre-factor $\dd\bar{K}/\dd \chi \propto \int \dd^2\bl W_S(\bl)^2\dd C_\ell^{\rm kSZ}/\dd \chi$, which can be expressed as:
\begin{eqnarray}
    \frac{\dd\bar{K}}{\dd \chi}\Bigg|_{\chi_*} = H(z_*)Q_*\VEV{v_{r,*}^2}\int \frac{\dd^2\bl}{(2\pi)^2} W_S^2(\ell) \Pion(\ell/\chi_*, z_*)\,,\nonumber\\
    \label{eq: dKbar_dz}
\end{eqnarray}
where the limits of the integral  $[\ell_{\rm min}, \ell_{\rm max}]$ determine the band-power of $K(\bn)$ construction, and we have defined $Q_* \equiv Q(z_*)$ and $\VEV{v_{r,*}^2} \equiv\VEV{v_r(z)^2}$ for ease of notation.

Although the $K(\bn)$ field is a line-of-sight integrated quantity, the specific contribution induced
within the 3D box at redshift $z_*$ can be isolated given a template for the $\eta(\bn, z)$ field within the desired comoving volume. 
Specifically, given a 3D tracer of $v_r(\bn,\,z_*)$, one can obtain a tomographic measurement of the cross-spectrum:
\begin{eqnarray}
    P_{K\eta}(L/\chi_*, z_*) = \frac{\dd\bar{K}}{\dd \chi}\Bigg|_{\chi_*}\frac{P_{\eta\eta}^\perp(L/\chi_*, z_*)}{\chi_*^2}\,,
    \label{eq: cross_spectrum}
\end{eqnarray}
where $\VEV{K(\bL)\eta(\bk)}_{z_*} \equiv P_{K\eta}(k, z_*) (2\pi)^3\delta^3(\bL/\chi_* + \bk)$ is the cross-correlation power spectrum at redshift $z_*$, and $P_{\eta\eta}^{\perp}(k, z)$ is the power spectrum of the $\eta(\bn,z)$ field, evaluated at wavenumber $\bk$ perpendicular to the line of sight [Eq.~\ref{eq: p_eta_perp}]. 
Here, we have once again leveraged the separate scales that source the kSZ effect---correlating the large-scale tracer of $v_r(\bn,z)^2$ with the $K(\bn)$ field results in a power-spectrum where the shape is entirely characterized by $P_{\eta\eta}^\perp(k)$ and the dependence on $P_{ee}^\ion(k, z_*)$ is embedded in the \textit{amplitude} of the cross-spectrum.
Therefore, given a fixed cosmological model, the above measurement can be used to place constraints on $P_{ee}^{\rm ion}(k, z_*)$ within a fixed bin $k\in[\ell_{\rm min}/\chi_*, \ell_{\rm max}/\chi_*]$. 

Since the galaxy-density field $\delta_g(\bk)$ is a tracer of the linear matter-overdensity field $\delta_m^{\rm lin}(\bk,z)$, galaxy-survey data can be used to reconstruct $v_r(\bn,z)$ via the linear-theory continuity equation [Eq.~\eqref{eq: galaxy_density_to_velocity}]. 
Therefore, galaxies are ideal for reconstructing $\eta(\bn,z)$ templates, enabling tomographic measurements of the ionized-electron distribution sourcing the kSZ signal.
Reference~\cite{AnilKumar:2025tbe} first proposed using galaxies to reconstruct the $\hat{\eta}(\bn)$ field, using them as the 3D tracer of $P_{K\eta}(L/\chi_*,z)$ to probe reionization~\footnote{See Ref.~\citep{AnilKumar:2025tbe} for detailed derivations of the $K(\bn)$- and $\eta(\bn)$-field signal and noise  reconstructions. See also Ref.~\citep{Caliskan:2023yov} for similar analysis using the fluctuations of optical depth as a 3-dimensional probe of reionization.}. In this work, we adopt the same strategy to instead characterize the low-redshift $P_{ee}^{\rm ion}(k, z_*)$. 

It is important to note that the proposed cross-correlation, although derived above in terms of $K(\bn)$ and $\eta(\bn,z)$, can be re-expressed as a cross-correlation between optimally filtered $T^2(\bn)$ and $\delta_g^2(\bk)$ (see App.~\ref{sec:the_3D_peculiar_velocity_squared_field} for details on $\eta$-reconstruction). 
The signal is, therefore, mathematically equivalent to a $\VEV{TT\delta_g\delta_g}$ four-point statistic, and we will henceforth use the terms “cross-spectrum” and “trispectrum” interchangeably.

\textit{Characterizing the Measurement SNR---}We characterize the sensitivity of this approach to $P_{ee}^{\rm ion}(k,z)$ by first constructing an optimal estimator for the cross-spectrum:
\begin{eqnarray}
    \hat{\mathcal{E}} = \int\frac{\dd^3\bk}{(2\pi)^3} \frac{\dd^2\bL}{(2\pi)^2}\ W(\bk, \bL)\eta(\bk)K(\bL)(2\pi)^3\delta^3(\bk+\bL/\chi_*)\,,\nonumber \\ 
    \label{eq:Ebox_general}
\end{eqnarray}
where $W(\bk, \bL)$ represents a weighting function that minimizes the variance of the estimator, subject to the constraint that $\langle{\hat{\mathcal{E}}}\rangle = 1$ if the true cross-spectrum is given by $P_{K\eta}$ from Eq.~\eqref{eq: cross_spectrum}. Note that the $z_*$ dependence 
of some quantities has been suppressed for ease of notation. 

Solving for the weights in this constrained optimization problem yields the following result:
\begin{eqnarray}
    \hat{\mathcal{E}} = N^2_{K\eta}\int\frac{\dd^3\bk}{(2\pi)^3} \frac{\dd^2\bL}{(2\pi)^2}\ \frac{P_{K\eta}(L/\chi_*)}{\tilde{P}^\perp_{\eta\eta}(k)\tilde{C}_L^{KK}}\eta(\bk)K(\bL)\nonumber\\\times(2\pi)^3\delta^3(\bk+\bL/\chi_*)\,,
    \label{eq:Ebox_final}
\end{eqnarray}
where, $\tilde{P}^\perp_{\eta\eta}(k) \equiv P^\perp_{\eta\eta}(k) + N^\perp_{\eta\eta}(k)$ is the galaxy-reconstructed $\eta$-field power-spectrum and $\td{C}_L^{KK}\equiv N_L^{KK}$ is the observed $K$-field power spectrum under the null-hypothesis. Here, $N^\perp_{\eta\eta}$ [Eq.~\eqref{eq: eta_recon_noise}] and $N_L^{KK}$ [Eq.~\eqref{eq: N_L_KK_expression}] are the noise in $\eta$- and $K$-field reconstruction, respectively.

The variance of this estimator is then given by the quantity $N^2_{K\eta}$, which can be calculated as follows:
\begin{eqnarray}
    N_{K\eta}^{-2} =V_0\int\frac{\dd^3\bk}{(2\pi)^3} \frac{\dd^2\bL}{(2\pi)^2}\ \frac{P_{K\eta}(L/\chi_*)^2}{\td{P}_{\eta\eta}(k)\td{C}_L^{KK}} (2\pi)^3 \delta^3(\bk + \bL/\chi_*)\,,\nonumber\\
    \label{eq: estimator_variance}
\end{eqnarray}
where $V_0 \equiv L_0^3$ is the volume of 3D the box centered at $z_*$, corresponding to the volume of the galaxy survey. The total signal-to-noise of this trispectrum measurement is SNR\,$=N_{k\eta}^{-1}$. 
The optimal high-pass filter to reconstruct $K(\bn)$, that maximizes this SNR, is then $W_S^2(\ell) \propto P_{ee}^\ion(\ell/\chi_*, z_*)/(\td{C}_\ell^{TT})^2$, where $\td{C}_\ell^{TT}$ is the total, observed CMB power spectrum.  

To evaluate the constraining power of a given experiment configuration, characterized by galaxy measurements in $[z, z + \dd z]$, CMB measurements in $[\ell,\ell+\dd\ell]$, and sky-overlap area $\Omega$, we can define the \textit{differential} SNR as $\dd{\rm SNR}^2 = \Omega[G(\ell,z)P_{ee}^{\rm ion}(\ell/\chi(z), z)^2]\dd z\dd\ell$, where
\begin{align}
    G(z, \ell) \equiv\; & \frac{H(z)}{4\pi}
    \left[\frac{Q(z)\VEV{v_r^2(z)}}{\chi(z)}\right]^2
    \frac{\ell}{\left(\td{C}_{\ell}^{TT}\right)^2} \nonumber\\[4pt]
    & \phantom{G(z, \ell) \equiv\;} \times 
    \left[\int \frac{\dd^2\bL}{(2\pi)^2}
    \frac{P_{\eta\eta}^\perp(L/\chi(z))^2}
         {\td{P}_{\eta\eta}^\perp(L/\chi(z))}\right]\,.
    \label{eq: K_eta_final_G_ell_z}
\end{align}
The differential SNR can then be integrated over the entire $z$-range corresponding to the volume of the survey, and the observed CMB multipole range to obtain the total SNR of measuring the trispectrum.
Moreover, we can use this definition to also estimate the statistical uncertainty $\Delta P_{ee}^{\rm ion}(k,z)$ with which the fiducial $P_{ee}^{\rm ion}(k,z)$ can be constrained, within a fixed $k\in [k_{\rm min}, k_{\rm max}]$ and $z \in [z_{\rm min}, z_{\rm max}]$:
\begin{eqnarray}
    \Delta P_{ee}
  = \left[ \Omega \int_{z_{\rm min}}^{z_{\rm max}} \int_{k_{\rm min}}^{k_{\rm max}} \dd z \, \dd k \, \chi(z) \, G(\ell,z)_{\ell=k\chi(z)} \right]^{-1/2}\,,\nonumber\\
  \label{eq: err_bars_P_ee_ion}
\end{eqnarray}
where we have assumed that $P_{ee}^{\rm ion}(k,z)$ has a constant value over the bin width.
In summary, we cross-correlate two fields: a 2D field $K(\bn)$ which is quadratic in the CMB, and a 3D field $\eta(\bn,z)$ which is quadratic in a galaxy field. 
The cross-correlation $P_{K\eta}(L/\chi(z),z)$ is proportional to $P_{ee}^\ion(k,z)$ via Eqs.~\eqref{eq: dKbar_dz}~and~\eqref{eq: cross_spectrum}. 
Then, we can either define multiple $K$-fields corresponding to different kSZ angular scales, to measure $P_{ee}^\ion(k,z)$ as function of two variables $\{k, z\}$, or define a single $K$-field integrated over all scales, to obtain an estimator $\hat{\mathcal{E}}$ to detect $P_{ee}^\ion(k,z)$ with maximal SNR.

A brief description of $\eta(\bn,z)$ reconstruction from galaxy observations, and the total $K(\bn)$ auto-power spectrum $\td{C}_L^{KK}$ is presented in App.~\ref{sec:the_3D_peculiar_velocity_squared_field} and App.~\ref{sec:the_kSZ_trispectrum}, respectively, for reference. 
Moreover, an explicit proof of the equivalence between this $K\times\eta$ formulation and the full $\langle TT\delta_g\delta_g\rangle$ trispectrum treatment is also provided in App.~\ref{sec:the_TTgg_trispectrum_approach}.

\textit{Results \& Forecasts---}The forecasts in this paper are calculated for two separate baselines:\,(1) ACT DR6-like CMB data~\citep{ACT:2020gnv,ACT:2023kun,ACT:2025xdm} cross-correlated with DESI spectroscopic galaxies~\citep{DESI:2022xcl,DESI:2013agm,DESI:2016fyo,2019AJ....157..168D,Zhou:2023gji}, and (2) Simons Observatory (SO)-like CMB observations~\citep{SimonsObservatory:2018koc,2019BAAS...51g.147L} combined with the same DESI sample. 
For both CMB experiments, we assume that $\td{C}_\ell^{TT} = C_\ell^{TT} + C_{\ell}^{\rm kSZ} + N_\ell^{\rm white}$, where $C_\ell^{TT}$ is the lensed, primordial CMB and $N_\ell^{\rm white}$ is the instrumental white-noise power spectrum. 
Moreover, we assume that the spectroscopic sample of galaxy locations collected by DESI are available up to $z = 2$, with galaxy number counts interpolated from Tab.~2 of Ref.~\cite{Hotinli:2024tjb}. 
The linear galaxy bias $b_g(z)$ is linearly interpolated from the results presented in Ref.~\cite{Yuan:2023ezi}.
Additionally, we assume that measurement of the galaxy power-spectrum is only limited by shot noise $[n_{\rm gal}(z)]^{-1}$, where $n_{\rm gal}(z)$ is the number density of galaxies observed within the redshift bin $[z, z+\dd z]$.  
The impact of using higher-density photometric galaxy samples from an LSST-like survey is explored in App.~\ref{sec:additional_results}.
For both baselines, we assume that the sky-overlap fraction is $f_{\rm sky} = 0.2$.
All relevant experiment specifications are summarized in Tab.~\ref{tab: experiment_specs}. 

\begin{table}
\centering
\renewcommand{\arraystretch}{1.5} 
\begin{tabularx}{\linewidth}{C|C|C|C}
\hline
 \textnormal{\textbf{CMB Survey}} & \textnormal{$\bm{\theta}_{\rm FWHM}$ \textbf{[arcmin]}} & \textnormal{$\bm{\Delta}_T$ \textbf{[}$\bm{\mu}$\textbf{K-arcmin]}} & \textbf{SNR} $\bm{z \in [0.1, 2.0]}$ \\
\hline
\hline
ACT & 1.6 & 15.0 &  7.5\\
SO & 1.4 & 5.8 &  31.4\\
\hline
\hline
\end{tabularx}

\vspace{0.65em}

\begin{tabularx}{\linewidth}{C|C|C|C}
\hline
\hline
\textnormal{\textbf{DESI Spec.}} & \textnormal{$\bm{z \in [0.1, 0.4]}$} & \textnormal{$\bm{z \in [0.4, 0.8]}$} & \textnormal{$\bm{z \in [0.8, 1.3]}$} \\
\hline
\hline
$n_{\rm gal}$ $[{\rm Mpc}^{-3}]$ & $4.8\times10^{-4}$ & $2.5\times10^{-4}$ & $1.2\times 10^{-4}$ \\
$b_g$ & $1.71$ & $1.97$ & 2.29 \\
\hline
\hline
\end{tabularx}
\caption{
\textit{Top:} Inputs to white-noise for the baseline CMB configurations. 
Parameter $\Delta_T$ is the amplitude of the noise and $\theta_{\rm FWHM}$ is the assumed resolution of the observed CMB map for ACT~\citep{ACT:2020gnv, ACT:2023kun, ACT:2025xdm, ACT:2023kun} and SO~\citep{SimonsObservatory:2018koc, 2019BAAS...51g.147L}. 
\textit{Bottom:} galaxy number-densities $n_{\rm gal}$ and biases $b_g$ assumed to characterize spectroscopic observations by DESI in the three equal-width (in $\chi$) redshift bins (for Fig.~\ref{fig:test_diff_baseline_comparison}). 
Total SNRs are computed by continuously interpolating the galaxy counts and biases in Tab.~2 of Ref.~\cite{Hotinli:2024tjb} and the results from Ref.~\cite{Yuan:2023ezi}, respectively.
For both baselines, we assume that $f_{\rm sky} = 0.2$.
} 
\label{tab: experiment_specs}
\end{table}

\begin{figure*}
    \centering
    \includegraphics[width=\linewidth]{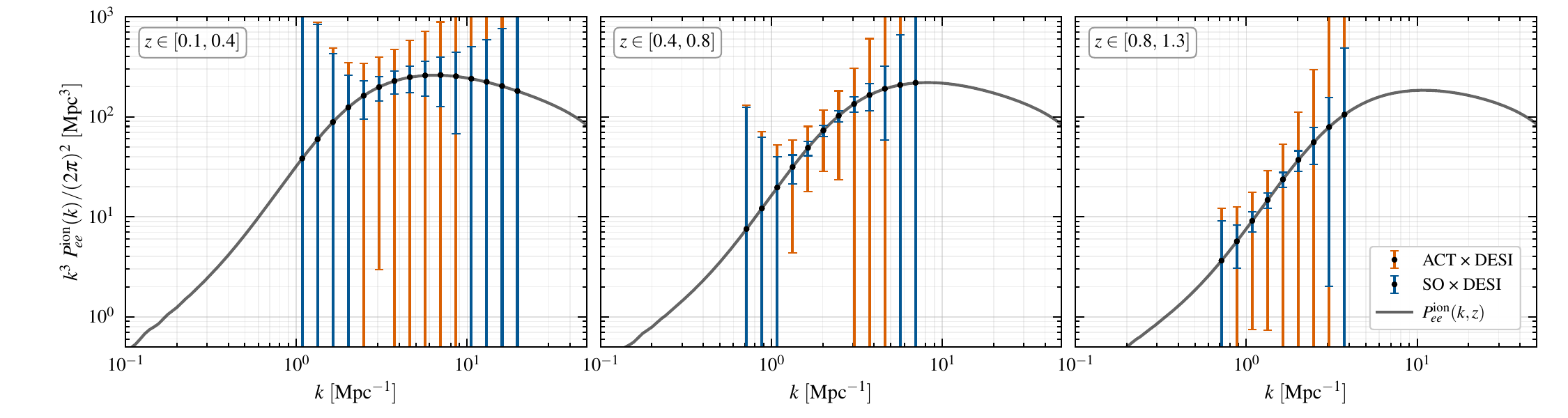}
    \caption{Forecasted errors on the reconstructed $P_{ee}^{\rm ion}(k,z)$ in three  $z$-bins of equal comoving width, with increasing redshift from left to right. Errors are computed using Eq.~\eqref{eq: err_bars_P_ee_ion}, assuming a fiducial $P_{ee}^{\rm ion}(k,z)$ following the `AGN' model from Ref.~\cite{Battaglia:2016xbi} on small scales (solid gray). Orange (blue) points denote the ACT$\times$DESI (SO$\times$DESI) baseline. Error bars are computed assuming a fixed $n_{\rm gal}$ and $b_g$ for each bin (Tab.~\ref{tab: experiment_specs}), and are shown only over the scales where uncertainties are minimum.
    }
    \label{fig:test_diff_baseline_comparison}
\end{figure*}

Forecasts presented in this section have been computed assuming that the small-scale distribution of electrons within halos can be characterized by the simulation-based `AGN' model in Ref.~\cite{Battaglia:2016xbi}. 
The impact of using an alternate model for $P_{ee}^\ion(k,z)$, with less power on small scales, is explored in App.~\ref{sec:additional_results}.
We ignore any contributions from reionization to the anisotropy captured by $P_{ee}^{\rm ion}(k,z)$ and $C_\ell^{\rm kSZ}$.
Further details regarding the computation of $P_{ee}^{\rm ion}(k,z)$ on small scales can be found in Sec.~II A and App. A of Ref.~\cite{AnilKumar:2025tbe}.

The third column of the top section of Tab.~\ref{tab: experiment_specs} displays the \textit{total} SNR of measuring the trispectrum, with the top (bottom) row corresponding to the result from the ACT$\times$DESI  (SO$\times$DESI) configuration. 
This quantity is computed by integrating the differential SNR [Eq.~\eqref{eq: K_eta_final_G_ell_z}] from $\ell_{\rm min} = 2500$ to $\ell_{\rm max} = 6000$, and $z_{\rm min} = 0.1$ to $z_{\rm max} = 2.0$ for both baselines.
The assumed $n_{\rm gal}(z)$ and $b_g(z)$ characterizing the DESI measurement are interpolated continuously across the redshift range from Tab.~2 of Ref.~\citep{Hotinli:2024tjb} and the results from Ref.~\citep{Yuan:2023ezi}, respectively. 
The total SNR as a function of $\ell_{\rm max}$ for both baselines is displayed in Fig.~\ref{fig:test_measurement_SNR} of App.~\ref{sec:additional_results} for reference.

Figure~\ref{fig:test_diff_baseline_comparison} shows the forecasted errors in reconstructing $P_{ee}^{\rm ion}(k,z)$ in three redshift bins of equal comoving width, ordered from lowest to highest $z$ from left to right.
The assumed fiducial $P_{ee}^{\rm ion}(k,z)$ is plotted in gray. 
Orange (blue) error bars correspond to the ACT$\times$DESI (SO$\times$DESI) baseline. 
The error-bars are computed using Eq.~\eqref{eq: err_bars_P_ee_ion}, assuming a fixed $n_{\rm gal}$ and $b_g$ across each bin (Tab.~\ref{tab: experiment_specs}), and are shown only for the range of scales where the forecasted uncertainties are smallest. 
These forecasts indicate that a CMB survey with specifications corresponding to ACT or SO can place competitive constraints on the morphology of ionized electron distributions on small scales, when cross-correlated with low number density spectroscopic galaxy samples from DESI, at instrument sensitivity. 
The lower resolution and higher noise-amplitude for ACT, however, limits the range of scales that can be accessed at high significance.
Furthermore, comparing the three panels indicates that smaller-scales of the power spectrum are accessible at lower redshifts.
This left-ward shift of the $k$-band within which $P_{ee}^{\rm ion}(k,z)$ is measurable is a manifestation of the fact that we are using mode $\ell$ from the CMB measurement to reconstruct scale $k \equiv \ell/\chi(z)$ at redshift $z$. 
 

Foregrounds in the observed CMB data, as well as non-Gaussianity in the $T_S^2(\bn)$ map from residual tSZ, cosmic infrared background (CIB), or CMB lensing, are expected to further degrade the reconstruction of $P_{ee}^{\rm ion}(k,z)$. 
In realistic analyses, component-separated maps may incur additional noise penalties at the multipoles relevant to this estimator, potentially degrading the SNR by up to an order of magnitude~\citep{SimonsObservatory:2018koc}. 
However, the impact of foreground mitigation is closely tied to the specific cleaning strategy adopted, and could be partially alleviated through approaches like ``bias hardening''~\citep{SPT-3G:2024lko, MacCrann:2024ahs}, needlet-based ILC
cleaning~\citep{McCarthy:2023hpa, Delabrouille:2008qd}, targeted combinations of temperature maps or the use of galaxy-based templates to trace contaminating power. 
Moreover, uncertainties in the measured galaxy bias $b_g(z)$ may introduce additional systematics in the reconstruction of $\eta(\bk)$.
We leave a comprehensive treatment of foreground mitigation and galaxy-bias systematics, tailored to this estimator, to future work.
Finally, although $C_\ell^{\rm kSZ}$ is sensitive to the $P_{ee}^{\rm ion}(k,z)$, the $\VEV{K\eta}$ cross-spectrum has several key advantages. 
The trispectrum yields \textit{tomographic} measurements of $P_{ee}^\ion(k,z)$, whereas $C_\ell^{\rm kSZ}$ is a sum of contributions from both reionization and low-redshift electron clustering, that are difficult to disentangle. 
Separating $C_\ell^{\rm kSZ}$ from other CMB secondaries such as lensing and the CIB is challenging, while the trispectrum estimator, defined as a cross-correlation between a CMB-derived field $K$ and an LSS-derived field $\eta$, is expected to be more robust to these systematics.

\textit{Conclusions}---In this paper, we have developed a method for measuring the distribution of ionized electrons from a novel kSZ$\times$galaxy four-point estimator, and showed that this higher-order statistic provides direct access to the cosmic baryon distribution. 
In contrast to CMB$\times$galaxy stacking analyses that probe the galaxy-electron cross-spectrum, our method directly accesses the small-scale, host-independent electron distribution without relying on small-scale galaxy clustering models.
Our forecasts demonstrate that current CMB data from ACT may already possess the sensitivity for a first detection of the electron distribution with this method, while near-future improvements in the CMB measurements by the SO will substantially sharpen its constraining power. 
Beyond establishing feasibility, these results highlight the opportunity to transform this kSZ$\times$galaxy statistic into a robust probe of baryonic feedback processes that displace and redistribute gas in and around halos. 
Such measurements will complement other kSZ, tSZ, and X-ray observations, and supplement constraints from the matter and lensing power spectra, thereby advancing our understanding of baryonic feedback and enhancing the scientific return from large-scale structure surveys.

\acknowledgements 

We thank Simone Ferraro, Matthew Johnson 
for useful discussions. This work was supported at JHU by NSF Grant No.\ 2412361, NASA ATP Grant No.\ 80NSSC24K1226, and the Templeton Foundation. SCH was supported by the P.~J.~E.~Peebles Fellowship at Perimeter Institute.
KMS was supported by an NSERC Discovery Grant, by the Daniel Family Foundation, and by the Centre for the Universe at Perimeter Institute.
Research at Perimeter Institute is supported by the Government of Canada through Industry Canada and by the Province of Ontario through the Ministry of Research \& Innovation.

\appendix
\section{The 3D $\eta(\bn,z)$ Field}
\label{sec:the_3D_peculiar_velocity_squared_field}
In this section, we provide a short summary of how galaxy survey measurements can be used to reconstruct the 3D $\eta(\bn)$ field. 
This reconstruction was first proposed by Ref.~\cite{AnilKumar:2025tbe}, where a cross-correlation between $K(\bn)$ and $\eta(\bn)$ was explored as a probe of He reionization. 

In the linear regime, the power-spectrum of the $\eta(\bn, s)$ field can be expressed as follows:
\begin{equation}
\begin{split}
P_{\eta\eta}^{\perp}(k)\!=\!\frac{2}{\langle v_r(z)^2\rangle^2}\!\!\int\!\!&\frac{\dd^3\boldsymbol{k}'}{(2\pi)^3}\frac{(k_r')^2(k_r-k_r')^2}{(k')^2\,(|\bk-\bk'|)^2}\\
&\times\!P_{vv}(k', z)P_{vv}(|\bk-\bk'|,z)\,,
\end{split}
\label{eq: p_eta_perp}
\end{equation}
where $P_{vv}(k,z)$ large-scale velocity power-spectrum, and the above integral is computed at wavenumbers $\bk$ perpendicular to the line of sight (i.e., assuming $k_r = 0$). 
Although the integrand in the above equation is redshift-dependent, in the linear regime the above expression for $P_{\eta\eta}^{\perp}(k)$ results in a $z$-independent quantity. 

The above signal can be reconstructed using galaxy-survey data given the continuity-equation-based relation between the cosmological velocity field $\boldsymbol{v}(\bk, z)$ and the large-scale matter overdensity field $\delta_m(\bk, z)$:
\begin{eqnarray}
    \hatv(\bk,z)= \frac{f(z)H(z)}{k(1+z)b_g(z)}{\hat{\delta}_g(\boldsymbol{k},z)}\,.
    \label{eq: galaxy_density_to_velocity}
\end{eqnarray}
In the above equation, $\hatv(\bk,z)$ is the Fourier domain, large-scale velocity field specifically reconstructed from galaxy-survey data, $f(z)$ refers to the linear growth rate $\dd \ln G/\dd \ln a$, and $\hat{\delta}_g(\bk,z)$ is the \textit{observed} galaxy density field.
The authors of Ref.~\cite{AnilKumar:2025tbe} find that the noise in reconstructed $\eta$-field power-spectrum can be expressed as:
\begin{eqnarray}
    N_{\eta\eta}^\perp(k) = \frac{P_{\eta\eta}^\perp(k)^2}{Q(k)} - P^\perp_{\eta\eta}(k)\,,
    \label{eq: eta_recon_noise}
\end{eqnarray}
where 
\begin{equation}
\begin{aligned}
    Q(k) \equiv \frac{2}{\VEV{v_r^2}^2}\int \frac{\dd^3\bk'}{(2\pi)^3}
    \, &\left(\frac{k_r'}{k'}\frac{(k_r - k_r')}{|\bk - \bk'|} \right)^2\\ &\times \frac{[P_{vv}(k')P_{vv}(|\bk-\bk'|)]^2}{\tilde{P}_{vv}(k')\td{P}_{vv}(|\bk-\bk'|)}\,,
\end{aligned}
\end{equation}
evaluated at wavenumber $\bk$ perpendicular to the line of sight. 
In the above expression, $\tilde{P}_{vv}(k,z) \equiv P_{vv}(k,z) +N_{vv}(k,z)$ represents the galaxy-reconstructed velocity power spectrum with reconstruction noise given by:
\begin{eqnarray}
    N_{vv}(k,z) &= &\left[\frac{f(z)H(z)}{k(1+z)b_g(z)}\right]^2 \frac{1}{n_{\rm gal}(z)}.
\end{eqnarray}

Further details on the derivation of the $\eta$-field reconstruction noise can be found in Sec. III B of Ref.~\cite{AnilKumar:2025tbe}. Note that the reconstruction noise has been recast in the form presented in Eq.~\eqref{eq: eta_recon_noise} for compactness, using the relation $\td{P}_{\eta\eta}^{\perp}(k) = P_{\eta\eta}^\perp(k)^2/Q(k)$, where $\td{P}_{\eta\eta}^{\perp}(k)$ is the \textit{observed} power spectrum of the reconstructed $\eta$-field. 
This result can be derived using the definition of the estimator and the optimal weights derived in Sec. III B of Ref.~\cite{AnilKumar:2025tbe}.

\section{The 2D $K(\bn)$ Field}
\label{sec:the_kSZ_trispectrum}
In this section, we provide a short summary of the kSZ trispectrum statistic, initially introduced in Ref.~\cite{Smith:2016lnt}, within the context of characterizing the high-$z$ epoch of hydrogen reionization. 
We present an expression for the observed power spectrum of the $K(\bn) \equiv T_S^2(\bn)$ field, assuming that the connected part of this four-point statistic is solely sourced by the kSZ anisotropy. 
In doing so, we clarify all simplifying assumptions made in arriving at the differential SNR expression detailed in Eq.~\ref{eq: K_eta_final_G_ell_z}.

The line-of-sight integrated $K(\bn)$ field can be expressed as:
\begin{eqnarray}
    K(\bn) = \int \dd z \frac{\dd \bar{K}}{\dd z} \eta(\bn, z)\,,
    \label{eq: K_nhat_eta_model}
\end{eqnarray}
where 
\begin{eqnarray}
    \frac{\dd\bar{K}}{\dd z} &= &\int \frac{\dd^2\bl}{(2\pi)^2} W_{S}^2(\ell)\frac{\dd C_\ell^{\rm kSZ}}{\dd z}\,,\nonumber \\
    &= &Q(z)\VEV{v_r(z)^2}\int \frac{\dd^2 \bl}{(2\pi)^2}W_S^2(\ell)P_{ee}^\ion(\ell/ \chi, z)\,.\nonumber \\ 
\end{eqnarray} 
The above expression accounts for the all contributions to the $K(\bn)$ field along the line of sight, instead of isolating the signal from a specific slice of redshift space. 
Therefore, in writing Eqs.~\eqref{eq: K_field_expression} and \eqref{eq: dKbar_dz} of the main text, we have made the assumption that $\dd\bar{K}/\dd\chi$ varies slowly across the width of the box centered at $z_*$. 

Under the Limber approximation, the power spectrum of the $K(\bn)$ field can then be written as:
\begin{eqnarray}
    C_L^{KK} = \int \dd z \frac{H(z)}{\chi(z)^2}\left(\frac{\dd \bar{K}}{dz}\right)^2P_{\eta\eta}^\perp(k = L/\chi, z)\,,
    \label{eq: C_L_KK_model_expression}
\end{eqnarray}
Given the above expression for the connected part of the kSZ trispectrum, the noise in the above measurement is given by:
\begin{eqnarray}
    N_L^{KK} = 2\int \frac{\dd^2\bl}{(2\pi)^2}W^2_{S}(\ell)W^2_{S}(|\boldsymbol{L}-\bl|)\td{C}_\ell^{TT}\td{C}_{|\boldsymbol{L} - \bl|}^{TT}\,,\nonumber\\
    \label{eq: N_L_KK_expression}
\end{eqnarray}
corresponding to the disconnected part of the kSZ trispectrum, i.e., the value $\tilde{C}_L^{KK}$ would take if the observed $T_S^2(\bn)$ map was perfectly Gaussian. 

In our forecasts, since $C_L^{KK}$ only appears in the variance of the proposed cross-correlation estimator [Eq.~\eqref{eq: estimator_variance}], we work under the \textit{null hypothesis} where $\td{C}_L^{KK} \approx N_L^{KK}$. 
Furthermore, since we are leveraging the cross-correlation $P_{K\eta}(L)$ for $L \ll \ell$, the expression for the $\tilde{C}_L^{KK}$ power spectrum under the null hypothesis can be approximated as:
\begin{eqnarray}
    \td{C}_L^{KK} = 2\int \frac{\dd^2\bl}{(2\pi)^2}W^4_{S}(\ell)(\td{C}_\ell^{TT})^2\,.
    \label{eq:C_L_KK_approximation}
\end{eqnarray}
This simplified expression is therefore plugged into Eq.~\eqref{eq: estimator_variance} of the main text, to determine that $W_S^2(\ell) \propto P_{ee}^{\rm ion}(\ell/\chi, z)/(\td{C}_\ell^{TT})^2$ optimizes the cross-spectrum detection SNR, resulting in the final expression for $G(\ell, z)$ in Eq.~\eqref{eq: K_eta_final_G_ell_z} of the main text.

\section{Additional Results}
\label{sec:additional_results}
This section presents a summary of supplementary results assessing the sensitivity of our forecasts to the minimum angular scale probed in the CMB map, photometric redshift uncertainties in galaxy measurements, and the assumed ionized electron profile.

\begin{figure}
    \centering
    \includegraphics[width=\linewidth]{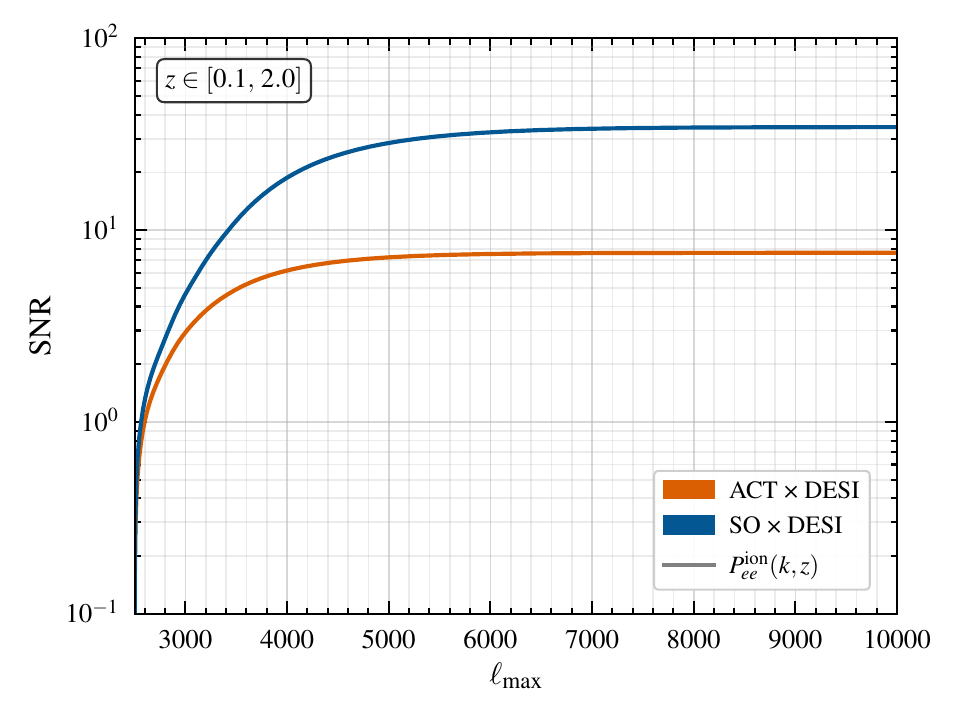}
    \caption{
    Forecasted total SNR of the trispectrum measurement as a function of the smallest accessible CMB scale $\ell_{\rm max}$. SNR is obtained by integrating over $\ell \in [2500,\,\ell_{\rm max}]$ and $z \in [0.1,\,2.0]$, with $n_{\rm gal}(z)$ interpolated from Ref.\cite{Hotinli:2024tjb} and $b_g(z)$ from Ref.\cite{Yuan:2023ezi}. Results are shown for $P_{ee}^{\rm ion}$ assuming the simulation-based model in Ref.~\cite{Battaglia:2016xbi} on small scales. Blue (Orange) curves indicate the SO$\times$DESI (ACT$\times$DESI) baseline.
    }
    \label{fig:test_measurement_SNR}
\end{figure}

Figure~\ref{fig:test_measurement_SNR} displays our forecasted total SNR of the trispectrum measurement for both the experiment configurations detailed in the main text, as a function of the smallest scale accessible in the CMB survey $\ell_{\rm max}$. 
The total SNR is computed by integrating over the scales $\ell \in [2500,\, \ell_{\rm max}]$ and redshifts $z \in [0.1,\, 2.0]$. 
For this calculation, the galaxy number density is interpolated from Tab. 2 of Ref.~\cite{Hotinli:2024tjb}, and the linear galaxy bias is interpolated from the results of Ref.~\cite{Yuan:2023ezi}. 
The results are computed assuming that the small-scale distribution of electrons within halos can be characterized by the simulation-based model in Ref.~\cite{Battaglia:2016xbi}. 
Blue curves correspond to the SO$\times$DESI baseline, while the orange curves represent the ACT$\times$DESI configuration. 

\begin{table}
\centering
\renewcommand{\arraystretch}{1.5} 
\begin{tabularx}{\linewidth}{C|C|C|C}
\hline
\hline
\textnormal{\textbf{LSST Spec.}} & \textnormal{$\bm{z \in [0.1, 0.4]}$} & \textnormal{$\bm{z \in [0.4, 0.8]}$} & \textnormal{$\bm{z \in [0.8, 1.3]}$} \\
\hline
\hline
$n_{\rm gal}$ $[{\rm Mpc}^{-3}]$ & $5.4\times10^{-2}$ & $2.8\times10^{-2}$ & $1.1\times 10^{-2}$ \\
$b_g$ & $1.04$ & $1.28$ & $1.59$ \\
$\sigma_z$ & $0.04$ & $0.05$ & $0.06$ \\
\hline
\hline
\end{tabularx}
\caption{
Galaxy number-densities $n_{\rm gal}$, biases $b_g$ and photometric redshift errors assumed to characterize observations by LSST in the three equal-width (in $\chi$) redshift bins. 
Forecasted errors in reconstructing $P_{ee}^{\rm ion}(k,z)$ are presented for these discrete bins in Fig.~\ref{fig:test_diff_gal_baseline_comparison}. 
For the LSST-based forecasts, we continue to assume that $f_{\rm sky} = 0.2$} 
\label{tab: LSST_specs}
\end{table}

To predict the impact of using a photometric galaxy-survey data set, we forecast uncertainties in the reconstruction of $P_{ee}^\ion(k,z)$ assuming LSST-like survey specifications characterizing the galaxy measurements. 
For these forecasts, we approximate  a fixed $n_{\rm gal}$ for each redshift bin assuming specifications matching the LSST “gold sample”:
\begin{eqnarray}
    n_\text{gal}(z) = n_0\left[\frac{z}{z_0}\right]^2\frac{\exp(-z/z_0)}{{2z_0}}\,,
\end{eqnarray}
where $n_0=40~\text{arcmin}^{-2}$, and $z_0=0.3$. 
Moreover, we assume that the galaxy bias also takes a fixed value in each bin, approximated using $b_g(z)=0.95(1+z)$.
Photometric (photo-$z$) redshift errors are characterized by parameter $\sigma_z(z)=0.03(1 + z)$, which we also assume takes a fixed value within a given bin. 
The effects of photometric measurements are accounted for by replacing the the velocity reconstruction noise $N_{vv}(k,z)$ with $W_{\sigma_z}(k_r, z)^{-2}N_{vv}(k,z)$, where
\begin{eqnarray}
    W_{\sigma_z}(k_r, z) = \exp\left(\frac{-\sigma_{z}^2}{2H(z)^2}k_r^2\right)\,.
\end{eqnarray}
All LSST survey parameters assumed for the forecasts that follow are detailed in Tab.~\ref{tab: LSST_specs}.

\begin{figure*}
    \centering
    \includegraphics[width=\linewidth]{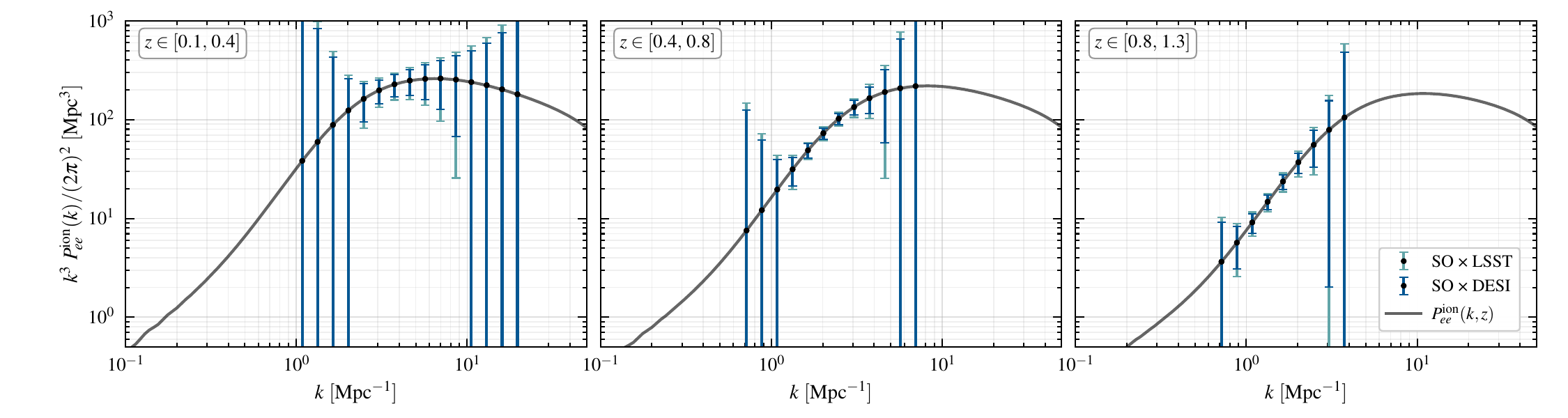}
    \caption{
    Forecasted errors on the reconstruction of $P_{ee}^{\rm ion}(k,z)$ in three redshift bins of equal comoving width, ordered from lowest to highest $z$ from left to right. 
    The fiducial $P_{ee}^{\rm ion}(k,z)$ is computed using the `AGN' model from Ref.~\cite{Battaglia:2016xbi}. 
    All error bars assume CMB data from an SO-like telescope (top of Tab.~\ref{tab: experiment_specs} of the main text). 
    Light-blue error bars show forecasts for the SO$\times$LSST baseline, while dark-blue error bars reproduce the previously forecasted results for the SO$\times$DESI baseline on the same grid for comparison. Experiment specifications assumed for DESI and LSST can be found in Tab.~\ref{tab: experiment_specs} of the main text and Tab.~\ref{tab: LSST_specs}, respectively.
    }
    \label{fig:test_diff_gal_baseline_comparison}
\end{figure*}

Figure~\ref{fig:test_diff_gal_baseline_comparison} finally shows the forecasted errors in reconstructing $P_{ee}^\ion(k,z)$ in three redshift bins of equal comoving width (matching those used in Fig.~\ref{fig:test_diff_baseline_comparison}), ordered from lowest to highest $z$ from left to right. 
The fiducial $P_{ee}^\ion(k,z)$ is computed assuming that the small-scale distribution of electrons the `AGN' profile from Ref.~\citep{Battaglia:2016xbi}. 
Both the sets of displayed error bars are computed assuming that the CMB data-set is obtained from an SO-like telescope (top of Tab.~\ref{tab: experiment_specs} of the main text). 
The light-blue error bars correspond to forecasts made assuming data from the SO$\times$LSST baseline. 
The previously forecasted measurement errors from the SO$\times$DESI baseline and re-produced on the same grid in dark blue for comparison. 

Interestingly, despite the significantly higher number-density of observed galaxies from an LSST-like survey, the forecasted errors across both the displayed baselines in Fig.~\ref{fig:test_diff_gal_baseline_comparison} are similar. 
This indicates that reconstruction of the $\eta(\bn,z)$-field via galaxy survey measurements is largely limited by cosmic variance, i.e., an increased number-density of observed galaxies (corresponding to lower shot noise in the observed galaxy power spectrum) does not significantly improve the sensitivity of this estimator to the small-scale distribution of ionized gas.
Moreover, the slight degradation in measurement precision from the SO$\times$LSST baseline, relative to SO$\times$DESI results, can be attributed to photo-$z$ errors.
This subdued impact of photo-$z$ errors is a manifestation of the fact that this estimator only relies on galaxy measurements for \textit{large-scale} velocity reconstruction and therefore does not require precise mapping of galaxy distributions on small scales. 

Finally, to demonstrate the dependence of our forecasts on the small-scale distribution of electrons (1-halo scale), we present results for an electron clustering model that predicts \textit{lower} power in the $k\gtrsim 1\ {\rm Mpc}^{-1}$ regime. 
Specifically, we adopt the `$W_e(k)\times$ NFW' model, an approximation of the small-scale electron distribution obtained by multiplying the non-linear matter power spectrum (obtained using an NFW model on
sub-halo scales) by a window function $W_e(k,z)$~\footnote{See App. A of Ref~\cite{AnilKumar:2025tbe} for the explicit functional form, and comparison to the assumed AGN profile and the NFW profile.}. Reference~\cite{Smith:2016lnt} explains that this fitting-function results in a
late-time kSZ power spectrum that approximately agrees
with the `cooling+star-formation' model from Ref.~\cite{Shaw:2011sy}.

For an ACT$\times$DESI configuration, assuming the experiment specifications detailed in Tab.~\ref{tab: experiment_specs}, $\ell_{\rm max} = 6000$, and that galaxy survey data from DESI extends to $z\sim 2$, we find a total SNR [Eq.~\eqref{eq: G_func_TTgg}] of trispectrum measurement to be $\sim 2.1$. 
Similarly, for an SO$\times$DESI configuration, we find a total SNR of $\sim 9.9$. 
The significant decrease in forecasted SNRs can be attributed to the lower small-scale power of the $W_e(k)\times$NFW model for $P_{ee}^\ion(k,z)$, which also results in a kSZ signal with much lower amplitude.

\begin{figure*}
    \centering
    \includegraphics[width=\linewidth]{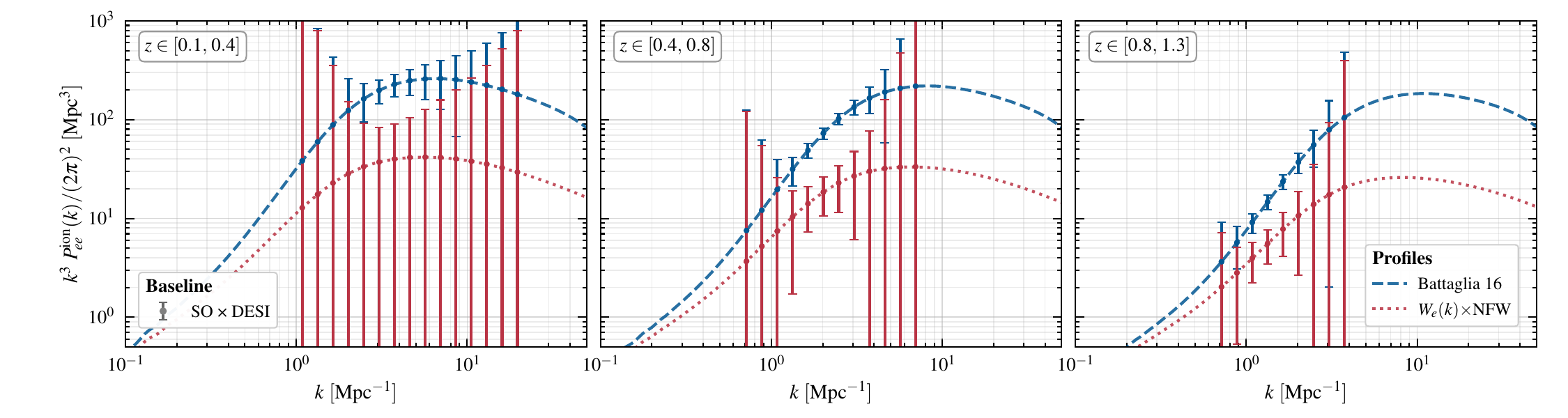}
    \caption{Forecasted errors on the reconstructed $P_{ee}^{\rm ion}(k,z)$ in three $z$-bins of equal comoving width, with increasing redshift from left to right (corresponding Fig.~\ref{fig:test_diff_baseline_comparison}). 
    Errors are computed using Eq.~\eqref{eq: err_bars_P_ee_ion} for the SO$\times$DESI baseline described in Tab.~\ref{tab: experiment_specs}, assuming fixed $n_{\rm gal}$ and $b_g$ within each bin. 
    The dotted red curve shows the fiducial `$W_e(k)\times$NFW' model, with red error bars denoting its forecasted uncertainties. 
    For comparison, the dashed blue curve shows the fiducial AGN feedback model from Ref.~\cite{Battaglia:2016xbi} (``Battaglia 16''), with corresponding forecasted uncertainties reproduced (Fig.~\ref{fig:test_diff_baseline_comparison}) in blue.
    Error bars are displayed only over the scales where the projected uncertainties are minimized.}
    \label{fig:test_diff_profile_comparison}
\end{figure*}

Figure~\ref{fig:test_diff_profile_comparison} shows forecasted errors in reconstructing $P_{ee}^\ion(k,z)$ in three redshift bins of equal comoving width, ordered from lowest to highest $z$ from left to right (corresponding exactly to the bins in Fig.~\ref{fig:test_diff_baseline_comparison}). 
The assumed fiducial `$W_e(k)\times$ NFW' model for $P_{ee}^\ion(k,z)$ is plotted as the dotted red curve, with red error bars computed using Eq.~\eqref{eq: err_bars_P_ee_ion}.
Forecasts are computed for the SO specifications detailed in top half of Tab~\ref{tab: experiment_specs}, assuming that the CMB measurements are combined with DESI galaxy survey data with fixed $n_{\rm gal}$ and $b_g$ across each bin, as indicated by the bottom half of Tab~\ref{tab: experiment_specs}.
For comparison, we also show in blue the forecasted uncertainties for the AGN feedback model under the same experimental configuration (Fig.~\ref{fig:test_diff_baseline_comparison}).
The associated fiducial prediction for $P_{ee}^\ion(k,z)$ is plotted as the dashed blue curve (``Battaglia 16").
As indicated by the SNR forecasts, the lower small-scale power in the  `$W_e(k)\times$ NFW' model makes the amplitude in each $k$-bin difficult to constrain. 
It is, however, important to note that the size of the error bars across the two models are roughly similar, because the choice of electron profile only impacts the kSZ contribution to $\td{C}_{\ell}^{TT}$ in Eq.~\eqref{eq: err_bars_P_ee_ion}.

\section{The $\VEV{TT\delta_g\delta_g}$ Trispectrum Approach}
\label{sec:the_TTgg_trispectrum_approach}
Earlier in this work, we derive the optimal estimator for measuring $P_{ee}^{\rm ion}(k,z)$ using the $K(\bn)$ and $\eta(\bn, z)$ fields. 
We adopt this approach for its compactness and its more intuitive interpretation, specifically within the context of a wealth of existing literature leveraging the kSZ trispectrum with the goal of probing the epochs of reionization~\citep{SPT-3G:2024lko, MacCrann:2024ahs,Chen:2022lhr,Ferraro:2018izc,Smith:2016lnt,Alvarez:2020gvl}. 
However, it is important to note that the optimal estimator for the proposed $P_{K\eta}$ cross-spectrum is mathematically equivalent to the optimal trispectrum estimator of type $\VEV{TT\delta_g\delta_g}$. 
In this section, we present the optimal trispectrum estimator, approaching the derivation through an explicit construction of the $\VEV{TT\delta_g\delta_g}$ four-point function. 

For the derivation that follows, we adopt the following Fourier conventions for 3D fields:
\begin{eqnarray}
    f(\br) = \int\frac{\dd^3\bk}{(2\pi)^3}f(\bk)e^{i\bk\cdot\br} \\ 
    f(\bk) = \int\dd^3 \br\, f(\br)e^{-i\bk\cdot\br}\,,
\end{eqnarray}
with similar conventions for 2D fields $T(\bn) \leftrightarrow T(\bl)$.

\subsection{Signal in 3D Box Formalism}
The real-space temperature anisotropy sourced by the kSZ effect within a 3D box centered at redshift $z_*$ can be expressed as:
\begin{eqnarray}
    T_{\rm kSZ}(\bn) = R_*\int_0^{L_0} \dd r\, \delta_e(\chi_*\bn,\  r\hat{\br})v_r(\chi_*\bn,\  r\hat{\br})\,,
\end{eqnarray}
where $\delta_e(\br)$ is the free-electron overdensity field, and $R_* \equiv R(z_*)$ is given by the following function:
\begin{eqnarray}
    R(z) \equiv T_{\rm CMB}H(z)\frac{\dd\bar{\tau}}{\dd z}e^{-\bar{\tau}(z)}\,,
\end{eqnarray}
evaluated at $z_*$. Moreover, we calculate the average optical depth to redshift $z$ using the following expression:
\begin{eqnarray}
    \bar{\tau}(z) = \sigma_T n_{e,0}\int_0^{z} \frac{\dd z' (1+z')^2}{H(z')}\bar{x}_e(z)\,,
    \label{eq: tau_bar_eq}
\end{eqnarray}
where $\sigma_T$ is the Thomson scatter cross-section, $n_{e,0}$ is the present number density of electrons, and $\bar{x}_e(z)$ is the average free-electron fraction at redshift $z$. Note that for the forecasts presented in this paper, we ignore any contributions to the kSZ signal sourced by the reionization, i.e., we set $\bar{x}_e = 1.0$ and only account for the kSZ signal induced at late times ($z \lesssim 5$), after the Universe is largely ionized.

Taking the Fourier transform of the above real-space map results in the following expression for the kSZ signal from $z_*$:
\begin{eqnarray}
    T(\boldsymbol{\ell}) = \frac{R_*}{\chi_*^2} \int \frac{\dd^3\boldsymbol{q}}{(2\pi)^3} & &\frac{\dd^3\boldsymbol{q}'}{(2\pi)^3} 
    \, \delta_e(\boldsymbol{q})\, v_r(\boldsymbol{q}')\nonumber \\
    & &\times (2\pi)^3 \delta^{(3)}\left(\boldsymbol{q} + \boldsymbol{q}' - \boldsymbol{\ell}/\chi_* \right)\,.
\end{eqnarray}
Given this expression, the four-point function can most-generally be written as:
\begin{widetext}
\begin{eqnarray}
    \VEV{T(\bl_1)T(\bl_2)\delta_g(\bk_1)\delta_g(\bk_2)} = \frac{R_*^2}{\chi_*^4} \int \frac{\dd^3\bq_1}{(2\pi)^3} \frac{\dd^3\bq_1'}{(2\pi)^3} & &\frac{\dd^3\bq_2}{(2\pi)^3}\frac{\dd^3\bq_2'}{(2\pi)^3}  \VEV{\delta_e(\bq_1')v_r(\bq_1)\delta_e(\bq_2')v_r(\bq_2)\delta_g(\bk_1)\delta_g(\bk_2)} \nonumber \\  & &\times(2\pi)^3 \delta^3(\bq_1 + \bq_1' - \bl_1/\chi_*)  \times (2\pi)^3 \delta^3(\bq_2 +\bq_2' - \bl_2/\chi_*)\,,
\end{eqnarray}
\end{widetext}
where $\delta_g(\bk)$ is the Fourier-domain, \textit{true} galaxy-overdensity field.
Under the null hypothesis, where the temperature squared map is Gaussian and uncorrelated with the galaxy-density field squared, the trispectrum above reduces to its disconnected part $\VEV{T(\bl_1)T(\bl_2)}\VEV{\delta_g(\bk_1)\delta_g(\bk_2)}$. 
The `signal' of interest is captured by the connected part of the above statistic $\VEV{T(\bl_1)T(\bl_2)\delta_g(\bk_1)\delta_g(\bk_2)}_c$. 
To simplify this calculation, we leverage the fact $\{\ell_1,\,\ell_2,\,q_1',\,q_2'\} \gg \{k_1,\,k_2,\,q_1,\,q_2\}$, i.e, we are cross-correlating the large scale variations in the locally-measured (small-scale) kSZ power with the long-wavelength galaxy density field. 
In this limit, we assume that the small-wavelength modes of $\delta_e(\bn)$ are uncorrelated with the long-wavelength modes of $\delta_g(\bn)$. 
Under these conditions, the connected part of the trispectrum can be simplified to the following form:
\begin{widetext}
    \begin{eqnarray}
        \VEV{T(\bl_1)T(\bl_2)\delta_g(\bk_1)\delta_g(\bk_2)}_c &= &\frac{R_*^2}{\chi_*^4}\frac{k_{1,r}k_{2,r}}{k_1k_2}[P_{ee}(|\bl_1/\chi_* - \bk_1|)P_{gv}(k_1)P_{gv}(k_2) + \bk_1 \leftrightarrow\bk_2] (2\pi)^3\delta^3(\bk_1 + \bk_2 + \bl_1/\chi_*  + \bl_2/\chi_*)\nonumber\\
        &= &\frac{2R_*^2}{\chi_*^4}\frac{k_{1,r}k_{2,r}}{k_1k_2}P_{ee}(\ell_1/\chi_*)P_{gv}(k_1)P_{gv}(k_2) (2\pi)^3\delta^3(\bk_1 + \bk_2 + \bl_1/\chi_*  + \bl_2/\chi_*)\,,\nonumber \\
        &\equiv &{\rm Tri}(\bl_1,\,\bl_2,\,\bk_1,\,\bk_2)(2\pi)^3\delta^3(\bk_1 + \bk_2 + \bl_1/\chi_*  + \bl_2/\chi_*)\,,
        \label{eq: TTgg_trispectrum}
    \end{eqnarray}
\end{widetext}
where the simplification in the second line arises from the fact that $\{k_1,\,k_2\} \ll \{ \ell_1,\, \ell_2 \}$.

\subsection{Optimal Estimator and Measurement SNR}
In line with the derivation presented within the main text of this work, we first derive its optimal estimator. 
Most generally, the optimal estimator for the connected four-point function can be written as:
\begin{widetext}
    \begin{eqnarray}
        \langle \hat{\mathcal{E}}\rangle = \int \ellInt{1}\ellInt{2}\kInt{1}\kInt{2} \left[T(\bl_1)T(\bl_2)\delta_g(\bk_1)\delta_g(\bk_2) - \td{C}_{\ell_1}^{TT}\td{P}_{gg}(k_1)\DdeltaS(\bl_1+\bl_2)\DdeltaC(\bk_1+\bk_2)\right] 
        \nonumber  \\ 
        \times  W(\ell_1, \ell_2, k_1, k_2) \DdeltaC(\bk_1+\bk_2+\bl_1/\chi_*+\bl_2/\chi_*)\,,
        \label{eq: TTgg_trispectrum_estimator}
    \end{eqnarray}
\end{widetext}
where $\td{P}_{gg}(k)$ is the total \textit{observed} galaxy power spectrum and $W(\ell_1, \ell_2, k_1, k_2)$ are optimal weights that minimize the noise in the connected trispectrum measurement. 
These weights must minimize the variance ${\rm Var}(\hat{\mathcal{E}})$, while subject to the constraint that $\hat{\mathcal{E}}$ is an unbiased estimator of the trispectrum amplitude [i.e., $\langle\hat{\mathcal{E}}\rangle = 1$ if the true trispectrum is given by Eq.~\eqref{eq: TTgg_trispectrum}].
In this derivation, we estimate the variance under the null hypothesis, i.e, assuming that $\delta_g$ and $T$ are uncorrelated Gaussian random fields. 
Using the Lagrange multipliers method, a straightforward calculation results in the following results for the optimal weights:
\begin{eqnarray}
    W(\bl_1,\,\bl_2,\,\bk_1,\,\bk_2) = \frac{\lambda\ {\rm Tri}(\bl_1,\,\bl_2,\,\bk_1,\,\bk_2)}{\td{C}_{\ell_1}^{TT}\td{C}_{\ell_2}^{TT}\td{P}_{gg}(k_1)\td{P}_{gg}(k_2)}\,.
\end{eqnarray}
In the above equation, the normalization constant $\lambda$ can be determined by plugging the weights into Eq.~\eqref{eq: TTgg_trispectrum_estimator} and imposing the constraint $\langle\hat{\mathcal{E}}\rangle = 1$. 
This results in the following expression for the estimator variance:
\begin{widetext}
    \begin{eqnarray}
        {\rm Var}(\hat{\mathcal{E}}) = \frac{4}{V}\lr{[}{\left(\frac{2R_*^2}{\chi_*^3}\right)^2\int \ellInt{}\kInt{1}\kInt{2}\left(\frac{k_{1,r}k_{2,r}}{k_1k_2}\right)^2\frac{[P_{ee}(\ell_1/\chi_*)P_{gv}(k_1)P_{gv}(k_2)]^2}{(\td{C}_{\ell}^{TT})^2\td{P}_{gg}(k_1)\td{P}_{gg}(k_2)}(2\pi)\delta(k_{1,r}+ k_{2,r})}{]}^{-1}\,,
    \end{eqnarray}
\end{widetext}
where we have not only plugged in the derived weights $W(\bl_1,\,\bl_2,\,\bk_1,\,\bk_2)$ with the appropriate expression for $\lambda$, but also simplified the result in the limit $\{k_1,\, k_2\} \ll \{\ell_1,\, \ell_2\}$.
Note that in the above expressions $k_{i, r}$ refers to the radial component of the wave-vector $\bk_i$.

The SNR of this measurement is simply given by ${\rm SNR} = {\rm Var}(\hat{\mathcal{E}})^{-1/2}$. 
Finally, the differential SNR, from a slice of redshift space $[z,\, z+\dd z]$, small-scale CMB multipole range $[\ell,\, \ell + \dd \ell]$, and observed sky area $\Omega$ can be written as $\dd{\rm SNR}^2 = \Omega[G(\ell,z)P_{ee}^{\rm ion}(\ell/\chi(z), z)^2]\dd z\dd\ell$, where 
\begin{widetext}
    \begin{eqnarray}
        G(\ell,z) = \frac{1}{2\pi H(z)}\lr{[}{\frac{R(z)^2}{\chi(z)^2}}{]}^2
        \lr{[}{\int \kInt{1}\kInt{2}
        \lr{(}{\frac{k_{1,r}k_{2,r}}{k_1k_2}}{)}^2\frac{[P_{gv}(k_1)P_{gv}(k_2)]^2}{\td{P}_{gg}(k_1)\td{P}_{gg}(k_2)}(2\pi)\delta(k_{1,r} +k_{2,r})}{]}
        \frac{\ell}{(\td{C}_{\ell}^{TT})^2}\,.
     \label{eq: G_func_TTgg}
    \end{eqnarray}
\end{widetext}
Given the results in Secs.~\ref{sec:the_3D_peculiar_velocity_squared_field} and ~\ref{sec:the_kSZ_trispectrum}, it can be shown that the functional form of $G(\ell,z)$ presented in the main text of this work [Eq.~\eqref{eq: K_eta_final_G_ell_z}] is mathematically equivalent to the expression derived above.

\bibliography{PRLdraft.bib}

\end{document}